\documentclass[reprint,superscriptaddress,amsmath, amssymb,aps,prd,notitlepage,longbibliography,floatfix,nofootinbib,onecolumn]{revtex4-1}

\usepackage{amsmath}
\usepackage{lipsum}
\allowdisplaybreaks[4]
\usepackage{cancel}
\usepackage{extarrows}
\usepackage{tensor}     
\usepackage{float}
\usepackage[caption = false]{subfig} 
\usepackage[final]{graphicx}   
\usepackage[
colorlinks=true,        
citecolor=blue,         
linkcolor=blue,         
urlcolor=blue           
]{hyperref}             
\usepackage{bm}         
\usepackage{xcolor}     
\usepackage{lipsum}
\usepackage{color}      
\usepackage[utf8]{inputenc} 
\usepackage[section]{placeins} 
\usepackage{appendix}
\usepackage{units}
\usepackage[capitalise]{cleveref}
\usepackage{units}

\usepackage{booktabs}
\usepackage{multirow}
\usepackage{import}
\usepackage{adjustbox}
\usepackage{url}
\newcommand{\nc}{\newcommand*}

\nc{\xbar}{\bar{x}}
\nc{\rhoeq}{\rho_{\mathrm{eq}}}
\nc{\zeq}{z_{\mathrm{eq}}}
\nc{\tla}{\tilde{\lambda}}
\nc{\bt}{\beta}
\nc{\dt}{\delta}
\nc{\Dt}{\Delta}
\nc{\vj}{\vec{j}}
\nc{\vl}{\vec{l}}
\nc{\hx}{\hat{x}}
\nc{\hy}{\hat{y}}
\nc{\bj}{\bm{j}}
\nc{\mJ}{\mathcal{J}}
\nc{\mP}{\mathcal{P}}
\nc{\Msun}{M_\odot}
\nc{\app}{\approx}
\nc{\av}[1]{\langle #1 \rangle}
\nc{\eq}[1]{Eq.~\eqref{#1}}
\nc{\al}{\alpha}
\nc{\Xstar}{X_{\ast}}
\nc{\fpbh}{f_{\mathrm{pbh}}}
\nc{\vth}{\vec{\theta}}
\nc{\vla}{\vec{\lambda}}
\nc{\vd}{\vec{d}}
\nc{\Mmin}{M_{\mathrm{min}}}
\nc{\rmd}{\mathrm{d}}
\nc{\mmin}{{m_{\mathrm{min}}}}
\nc{\mmax}{{m_{\mathrm{max}}}}
\nc{\mR}{\mathcal{R}}
\nc{\tmR}{\tilde{\mathcal{R}}}
\nc{\s}{\sigma}
\nc{\ogw}{\Omega_{\mathrm{GW}}}
\nc{\addref}{[\textcolor{red}{add ref}] }
\nc{\Om}{\Omega}
\nc{\gm}{\gamma}
\nc{\Gm}{\Gamma}
\nc{\gpcyr}{\mathrm{Gpc}^{-3}\,\mathrm{yr}^{-1}}
\nc{\Eq}[1]{Eq.~\eqref{#1}}
\nc{\Fig}[1]{Fig.~\ref{#1}}
\nc{\Table}[1]{Table~\ref{#1}}
\nc{\lvc}{LIGO/Virgo} 
\nc{\Sec}[1]{Sec.~\ref{#1}}
\nc{\eg}{\textit{e.g.~}}
\nc{\SNR}{\mathrm{SNR}}
\nc{\be}{\mathbf{\epsilon}}
\nc{\bn}{\mathbf{n}}
\nc{\bd}{\mathbf{d}}
\nc{\ba}{\mathbf{a}}
\nc{\eps}{\epsilon}
\nc{\bnu}{\mathbf{\nu}}
\nc{\mb}{\mathbf}
\nc{\bbt}{\mathbf{t}}
\nc{\bth}{\mathbf{\theta}}
\nc{\bep}{\mathbf{\epsilon}}
\nc{\uni}{\mathrm{U}}
\nc{\logu}{\operatorname{\mathrm{log-U}}}
\nc{\RN}{\mathrm{RN}}
\nc{\BN}{\mathrm{BN}}
\nc{\GN}{\mathrm{GN}}
\nc{\mcN}{\mathcal{N}}
\nc{\GWB}{\mathrm{GW}}
\nc{\yr}{\mathrm{yr}}
\nc{\Am}{\mathcal{A}}
\nc{\Dm}{\mathcal{D}}
\nc{\Hm}{\mathcal{H}}
\nc{\sovast}{Soviet Ast.}

\nc{\kmsmpc}{\mathrm{km\ s^{-1} Mpc^{-1}}}
\nc{\lcdm}{\Lambda\mathrm{CDM}}
\nc{\ev}{\mathrm{eV}}

\nc{\mrm}{\mathrm}
\nc{\BE}{B\scriptsize{AYES}\normalsize{E}\scriptsize{PHEM}\normalsize  }

\nc{\Ostgw}{\Omega_{\mathrm{GW}}^{\mathrm{ST}}}
\nc{\Ottgw}{\Omega_{\mathrm{GW}}^{\mathrm{TT}}}
\nc{\Ovlgw}{\Omega_{\mathrm{GW}}^{\mathrm{VL}}}
\nc{\Oslgw}{\Omega_{\mathrm{GW}}^{\mathrm{SL}}}
\nc{\cosxi}{\beta}

\nc{\gmPL}{\gamma_{\mathrm{PL}}}
\nc{\APL}{A_{\mathrm{PL}}}

\def\({\left(}
\def\){\right)}
\def\[{\left[}
\def\]{\right]}

\def\e{\begin{equation}}
\def\q{\end{equation}}
\def\m{\begin{eqnarray}}
\def\n{\end{eqnarray}}
\nc{\red}[1]{\textcolor{red}{#1}}

\begin{document}


\title{Reevaluating $H_0$ Tension with Non-\textit{Planck} CMB and DESI BAO Joint Analysis}

\author{Ye-Huang Pang}
\email{pangyehuang22@mails.ucas.ac.cn}
\affiliation{School of Fundamental Physics and Mathematical Sciences, Hangzhou Institute for Advanced Study, UCAS, Hangzhou 310024, China}
\affiliation{School of Physical Sciences,
    University of Chinese Academy of Sciences,
    No. 19A Yuquan Road, Beijing 100049, China}
\affiliation{CAS Key Laboratory of Theoretical Physics,
    Institute of Theoretical Physics, Chinese Academy of Sciences,Beijing 100190, China}
\author{Xue Zhang}
\email{corresponding author: zhangxue@yzu.edu.cn}
\affiliation{Center for Gravitation and Cosmology,
    College of Physical Science and Technology,
    Yangzhou University, Yangzhou 225009, China}
\author{Qing-Guo Huang}
\email{corresponding author: huangqg@itp.ac.cn}
\affiliation{School of Fundamental Physics and Mathematical Sciences, Hangzhou Institute for Advanced Study, UCAS, Hangzhou 310024, China}
\affiliation{School of Physical Sciences,
    University of Chinese Academy of Sciences,
    No. 19A Yuquan Road, Beijing 100049, China}
\affiliation{CAS Key Laboratory of Theoretical Physics,
    Institute of Theoretical Physics, Chinese Academy of Sciences,Beijing 100190, China}


\begin{abstract}
$H_0$ tension in the spatially flat $\Lambda\mathrm{CDM}$ model is reevaluated by employing three sets of non-\textit{Planck} CMB data, namely WMAP, WMAP+ACT, and WMAP+SPT, in conjunction with DESI BAO data and non-DESI BAO datasets including 6dFGS, SDSS DR7, and SDSS DR16. Our analysis yields $H_0 = 68.86\pm 0.68~\mathrm{km\ s^{-1} Mpc^{-1}}$ with WMAP+DESI BAO, $H_0 = 68.72\pm 0.51~\mathrm{km\ s^{-1} Mpc^{-1}}$ with WMAP+ACT+DESI BAO, and $H_0 = 68.62\pm 0.52~\mathrm{km\ s^{-1} Mpc^{-1}}$ with WMAP+SPT+DESI BAO. 
The results of non-\textit{Planck} CMB+DESI BAO exhibit a $3.4\sigma$, $3.7\sigma$, and $3.8\sigma$ tension with the SH0ES local measurement respectively which are around $1 \sigma$ lower in significance for the Hubble tension compared to \textit{Planck} CMB+DESI BAO.
Moreover, by combining DESI BAO data+non-\textit{Planck} CMB measurements, we obtain a more stringent constraint on the Hubble constant compared to non-DESI BAO data+non-\textit{Planck} CMB data, as well as reducing the significance of the Hubble tension. 
\end{abstract}
\maketitle

\section{Introduction}
Hubble tension, arising from the discrepancy in the measurement of $H_0$ between the cosmic microwave background (CMB) and the local distance ladder method, has been extensively discussed over an extended period.
The \textit{Planck} 2018 TT,TE,EE+lowE+lensing dataset provides a constraint of $H_0 = 67.27 \pm 0.60$ $\kmsmpc$ \cite{Planck:2018nkj}, while the Supernovae and $H_0$ for the Equation of State of dark energy (SH0ES) project reported $H_0 = 73.04 \pm 1.04 ~\kmsmpc$, measured with local distance ladder calibrated by Cepheid variables \cite{Riess:2021jrx}. There is a significant discrepancy at $4.8\sigma$ level between these two independent measurements of $H_0$.
The reason for this discrepancy are still bewildering. This situation will likely persist until we identify the systematic errors in specific measurements or uncover flaws in our current cosmological model, ultimately confirming a revised model through precise observations and consistency checks. For an overview of the Hubble tension and attempts to solve this problem, refer to Refs. \cite{Verde:2019ivm, DiValentino:2021izs, Abdalla:2022yfr, Kamionkowski:2022pkx}. 

The availability of various datasets allows us to perform constraints using different data combinations and assess their consistency, thereby helping us evaluate the current urgency of addressing the Hubble tension.
In addition to \textit{Planck} CMB data, Wilkinson Microwave Anisotropy Probe (WMAP) data \cite{WMAP:2006bqn} are also available for cosmological parameter inference. Ground based CMB observatories such as the Atacama Cosmology Telescope (ACT) \cite{ACTPol:2016kmo, ACT:2020gnv}, the South Pole Telescope (SPT) \cite{SPT:2017jdf, SPT-3G:2021eoc, SPT-3G:2022hvq}, Background Imaging of Cosmic Extragalactic Polarization and the Keck Array (BICEP/Keck) \cite{BICEP2:2015vut} also provide valuable complementary information.
The WMAP year-9 result yields a value of $H_0 = 70.0 \pm 2.2~\kmsmpc$ \cite{WMAP:2012nax}. The Hubble tension does not emerge due to the large mean value and error bar of $H_0$, which is the result of the limited number of multipoles provided by WMAP. However, when combined with ACT or SPT datasets \cite{ACT:2020gnv, Calderon:2023obf}, this lack of high multipole information is compensated for and leads to improved constraining power.

Incorporating CMB and BAO data can further enhance the precision of $H_0$ constraints, which has been extensively utilized to constrain $H_0$ and other cosmological parameters in numerous studies \cite{Zhang:2018air, DESI:2024lzq, Pogosian:2024ykm, Efstathiou:2024dvn}. Among these studies, the combination of Planck 2018+SDSS BAO yields a result of $H_0 = 67.66\pm 0.42~\kmsmpc$, $\Omega_m = 0.3111 \pm 0.0056$ \cite{Planck:2018vyg}, which differs from SH0ES by $4.8 \sigma$. 
Utilizing the latest BAO data from the first year of DESI observations, it has been reported that $H_0 = 67.97 \pm 0.38~\kmsmpc$ and $\Omega_m = 0.3069 \pm 0.0050$ with \textit{Planck} 2018+ACT lensing+DESI BAO \cite{DESI:2024lzq}, showing a tension of $4.6\sigma$ with the SH0ES measurement. 
The implications of DESI BAO on alleviating the $H_0$ tension have also been discussed in Refs. \cite{Giare:2024smz, Qu:2024lpx, Wang:2024dka, Seto:2024cgo, Lynch:2024hzh, Jiang:2024xnu, Escamilla:2024xmz, Shepelev:2024guu, RoyChoudhury:2024wri, Wang:2024tjd}. 
Additionally, differences between DESI BAO data and SDSS BAO data (or other non-DESI BAO) can lead to varying results \cite{Colgain:2024xqj, Chan-GyungPark:2024spk, Dinda:2024ktd, Ghosh:2024kyd}.

Since the constraints of $\lcdm$ model with non-\textit{Planck} data and DESI BAO have not been addressed in prior works, we aim to investigate these constraints and their implications for the $H_0$ tension. 
In this article, using these datasets we evaluate the Hubble constant within a spatially flat $\lcdm$ model. and present our findings. Additionally, we analyze the constraints from non-\textit{Planck} and non-DESI BAO for comparative purposes. 
In \Sec{data}, we describe the details of the data that we used. The results are presented in \Sec{result}. We discuss and summarize our findings in \Sec{summary}.

\section{Datasets and Methods}\label{data}

We perform our analysis using three different CMB datasets: WMAP alone, WMAP combined with ACT, and WMAP combined with SPT. These datasets are all of which are independent of \textit{Planck} observations.
For our analysis, we use 9-year observational results from WMAP, which include the TT power spectrum within multipole ranges of $2 < \ell < 1200$ and $24 < \ell < 800$ for the EE power spectrum \cite{WMAP:2012nax}. 
Additionally, we incorporate ACT data using the ACTPol DR4 likelihood implemented in \texttt{pyactlike}\footnote{\url{https://github.com/ACTCollaboration/pyactlike}} \cite{ACT:2020gnv}. The SPT data consist of a TT power spectrum within $750 < \ell < 3000$, as well as TE and EE spectra within  $300< \ell < 3000$ \cite{SPT-3G:2022hvq}. 
When combining WMAP with either dataset, we disregard any correlation between WMAP and ACT or between WMAP and SPT.

We include the utilized DESI BAO data \cite{DESI:2024uvr, DESI:2024lzq, DESI:2024mwx} and non-DESI BAO data in \Table{tab:bao}. 
The non-DESI BAO data include 6dFGS \cite{Beutler:2011hx}, SDSS DR7 MGS \cite{Ross:2014qpa} and SDSS DR16 data \cite{eBOSS:2020yzd}.
\begin{table}[!htbp]
    \centering
    \begin{tabular} { l  l  c  c}
        \hline
          & &\ $z_{\mathrm{eff}}$ \  &\ measured quantities and values\ \\
        \hline
        DESI BAO data
        &{BGS}
        &$0.30$
        &$D_V/r_d = 7.93\pm0.15 $\\
        
        \ 
        &{LRG}
        &$0.51$
        &$D_M/r_d = 13.62\pm 0.25,\ D_H/r_d=20.98\pm 0.61 $\\
        
        \ 
        &{LRG}
        &$0.71$
        &$D_M/r_d = 16.85\pm 0.32,\ D_H/r_d=20.08\pm 0.60 $\\
        
        \ 
        &{LRG+ELG}
        &$0.93$
        &$D_M/r_d = 21.71\pm 0.28,\ D_H/r_d=17.88\pm 0.35 $\\
        
        \ 
        &{ELG}
        &$1.32$
        &$D_M/r_d = 27.79\pm 0.69,\ D_H/r_d=13.82\pm 0.42 $\\
        
        \ 
        &{QSO}
        &$1.49$
        &$D_V/r_d = 26.07\pm0.67 $\\
        
        \ 
        &{Ly$\alpha$ QSO}
        &$2.33$
        &$D_M/r_d = 27.79\pm 0.69,\ D_H/r_d=13.82\pm 0.42 $\\
        \hline
        
        non-DESI BAO data
        &{6dFGS}
        &$0.106$
        &$r_d/D_V = 0.336\pm 0.015$\\
        
        \ 
        &{SDSS DR7 MGS}
        &$0.15$
        &$D_V/r_d= 4.47\pm 0.17$\\
        
        \ 
        &{SDSS DR16 LRG}
        &$0.70$
        &$D_M/r_d = 17.86\pm 0.33,\ D_H/r_d = 19.33\pm 0.53 $\\
        
        \ 
        &{SDSS DR16 ELG}
        &$0.85$
        &$D_V/r_d = 18.33^{+0.57}_{-0.62} $\\
        
        \ 
        &{SDSS DR16 QSO}
        &$1.48$
        &$D_M/r_d = 30.69\pm 0.80,\ D_H/r_d = 13.26\pm 0.55$\\
        
        \ 
        &{SDSS DR16 Ly$\alpha$}
        &$2.33$
        &$D_M/r_d = 37.6\pm 1.9,\ D_H/r_d = 8.93\pm 0.28$\\
        
        \ 
        &{SDSS DR16 Ly$\alpha$ QSO}
        &$2.33$
        &$D_M/r_d = 37.3\pm 1.7,\ D_H/r_d = 9.08\pm 0.34$\\
        \hline
    \end{tabular}
    \caption{DESI BAO data and non-DESI BAO data used in our analysis.}
    \label{tab:bao}
\end{table}
In all BAO measurment, the sound horizon at the baryon drag epoch, denoted as $r_d$, is defined as
\begin{equation}
        r_d = \int_{z_d}^{\infty} \dfrac{c_s(z)}{H(z)} \mathrm{d}z,
\end{equation}
where $c_s(z)$ represents the sound speed of the baryon-photon fluid. 
The transverse comoving distance $D_M(z)$ can be calculated by
\begin{equation}
    D_M(z) = \int_{0}^{z} \dfrac{c\ \mathrm{d} z^\prime}{H(z^\prime)},
\end{equation}
as well as
\begin{equation}
    D_H(z) = c/H(z).
\end{equation}
Additionally, the angle-averaged distance $D_V$ is given by
\begin{equation}
    D_V(z) = \left(z D_M(z)^2 D_H(z)\right)^{1/3}.
\end{equation}
%

Including supernova data can further enhance the constraint on cosmological parameters. However, the systematic errors in different supernova datasets may non-trivially affect the final results and complicate our discussion. Therefore, we do not include supernova data in our analysis.

We use the \texttt{class} package \cite{Blas:2011rf} for theoretical calculations and preform MCMC sampling with the \texttt{cobaya} package \cite{Torrado:2020dgo}. The chains that satisfy the Gelman-Rubin criteria \cite{Gelman:1992zz}, $R-1<0.01$, are considered to be converged. 

\section{Results}\label{result}

\begin{figure}[!htbp]
    \centering
        \includegraphics[width=0.5\textwidth]{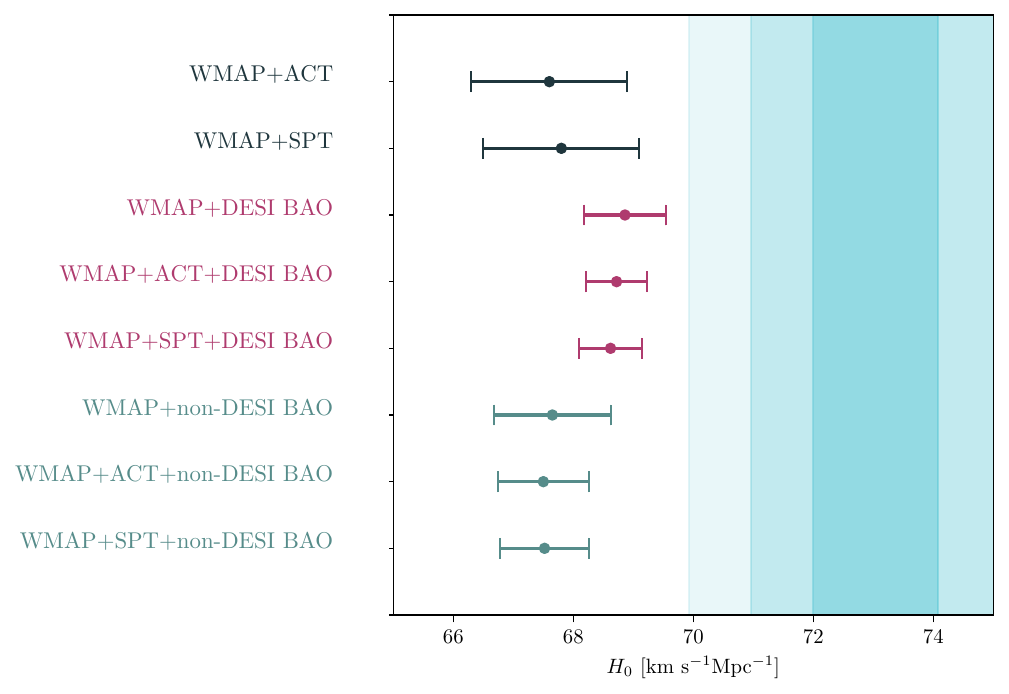}
        \includegraphics[width=0.333\textwidth]{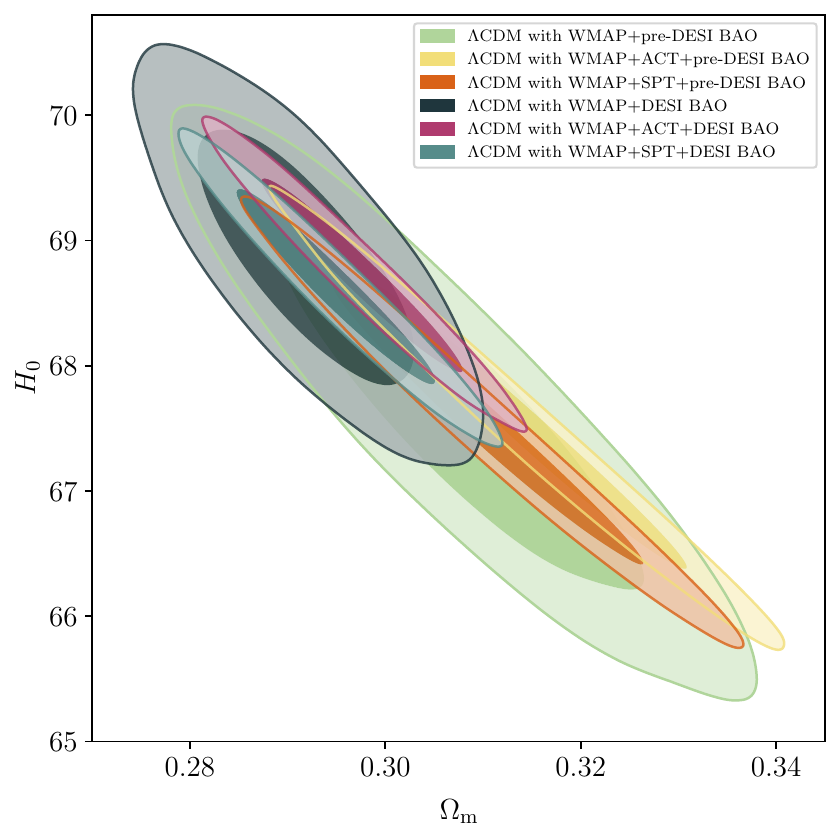}
    \caption{Left panel: $H_0$ constraints at the $1\sigma$ level, with the azure band representing the SH0ES $H_0$ constraint at the $1\sigma$, $2\sigma$ and $3\sigma$ levels. Right panel: 2-dimensional marginalized posterior distribution of $H_0$ and $\Omega_m$.}
    \label{fig:fig1}
\end{figure}

In \Fig{fig:fig1}, we present the error bar plots of $H_0$ at $ 1\sigma$ confidence levels (C.L.) and the contours of $H_0\text{-}\Omega_m$ plane. The corresponding mean values and $ 1\sigma$ C.L. of $\Omega_m$ and $H_0$ are listed in \Table{tab:table2}.
The results show that the CMB-only measurements from WMAP+ACT and WMAP+SPT are in $\sim 3\sigma$ tension with the value measured by SH0ES. However, it is important to note that the error bars in these measurements are relatively large compared to the constraints derived from CMB+BAO data. 


\begin{table}[!htbp]
    \centering
    \begin{tabular} { l  c  c  c}
        \hline
          &  \ $\Omega_m$ \  &\ $H_0 [\mathrm{km\  s^{-1} Mpc^{-1}}]$ \  &\ $H_0$ tension\ \\
        \hline

        {WMAP+ACT}
        & $0.314\pm 0.018            $
        & $67.6\pm 1.3               $
        & $3.3\sigma$\\

        {WMAP+SPT}
        & $0.307\pm 0.018            $
        & $67.8\pm 1.3               $
        & $3.1\sigma$\\
        \hline
        {WMAP+DESI BAO}
        & $0.2919\pm 0.0073          $
        & $68.86\pm 0.68             $
        & $3.4\sigma$\\

        {WMAP+ACT+DESI BAO}
        & $0.2976\pm 0.0067          $
        & $68.72\pm 0.51             $
        & $3.7\sigma$\\

        {WMAP+SPT+DESI BAO}
        & $0.2950\pm 0.0068          $
        & $68.62\pm 0.52             $
        & $3.8\sigma$
        \\
        \hline
        {WMAP+non-DESI BAO}
        & $0.307\pm 0.012$
        & $67.65\pm 0.97$
        & $3.8\sigma$
        \\

        {WMAP+ACT+non-DESI BAO}
        & $0.315\pm 0.011$
        & $67.50\pm 0.76$
        & $4.3\sigma$
        \\

        {WMAP+SPT+non-DESI BAO}
        & $0.310\pm 0.011    $
        & $67.52\pm 0.74             $
        & $4.3\sigma$
        \\
        \hline
    \end{tabular}
    \caption{The mean value and $1\sigma$ C.L. of $\Omega_m$ and $H_0$, constrained with various data combinations, and corresponding $H_0$ tension compared with SH0ES measurement of  $H_0$.}
    \label{tab:table2}
\end{table}

The joint analysis from BAO and CMB data helps to reduce the error of $H_0$. For the combination of WMAP+DESI BAO, we find $H_0 = 68.86\pm 0.68~\kmsmpc$, which shows a deviation from the SH0ES local distance ladder measurement at $3.4\sigma$ significance level. When using the joint constraint of WMAP+ACT+DESI BAO, we obtain $H_0 = 68.72\pm 0.51~\kmsmpc$ ($3.7\sigma$ tension with SH0ES). Similarly, for the combination of WMAP+SPT+DESI BAO, we find $H_0 = 68.62\pm 0.52~\kmsmpc$ ($3.8\sigma$ tension with SH0ES).
Furthermore, the combination of non-\textit{Planck} CMB+DESI BAO provides a tighter constraint on $H_0$ compared to the non-\textit{Planck} CMB+non-DESI BAO cases. This indicates a smaller deviation from the SH0ES result and a shift towards lower values in $\Omega_m$. The stronger constraints are also visually evident in the two-dimensional marginalized distribution of $H_0$ and $\Omega_m$, as depicted in \Fig{fig:fig1}.

As reported in Ref. \cite{DESI:2024lzq}, the combination of \textit{Planck}+ACT lensing+DESI BAO yields $H_0 = 67.97 \pm 0.38~\kmsmpc$, which exhibits a $4.6\sigma$ discrepancy with the result of SH0ES. However, when we replace the CMB data form \textit{Planck} with WMAP, WMAP+ACT and WMAP+SPT, we observe that the Hubble tension is reduced by $\sim 1 \sigma$. This suggest that \textit{Planck} data may contribute to exacerbating the $H_0$ tension.

\section{Summary and Discussion}\label{summary}

We use three distinct sets of non-\textit{Planck} CMB data, namely WMAP, WMAP+ACT and WMAP+SPT, in conjunction with DESI and non-DESI BAO data to assess the value of $H_0$ assuming a flat $\lcdm$ model.

The joint constraints of non-\textit{Planck} CMB+DESI BAO help alleviate Hubble tension. For the data combinations WMAP+DESI BAO, WMAP+ACT+DESI BAO, and WMAP+SPT+DESI BAO, we obtain $H_0 = 68.86\pm 0.68~ \kmsmpc$, $68.72\pm 0.51~\kmsmpc$, and $68.62\pm 0.52 ~\kmsmpc$ respectively. 
The corresponding tensions with the SH0ES local measurement for these data combinations are at significance levels of $3.4\sigma,\  3.7\sigma,\ 3.8\sigma$ respectively.
Our analysis suggests that the combinations of CMB data independent of \textit{Planck} with DESI BAO data reduce the significance of Hubble tension compared to the result of $H_0 = 67.97 \pm 0.38~ \kmsmpc$ constrained with \textit{Planck}+ACT lensing+DESI BAO. The substitution of non-\textit{Planck} datasets for \textit{Planck} data appears to be the key factor in reducing $H_0$ tension.

Additionally, we derived the constrains on $H_0$ with WMAP+non-DESI BAO, WMAP+ACT+non-DESI BAO and WMAP+SPT+non-DESI BAO, yielding values of $67.65\pm 0.97 ~\kmsmpc, 67.50\pm 0.76~\kmsmpc$, and $67.52\pm 0.74 ~\kmsmpc$ respectively. These  results exhibit a significant deviation from the value reported by SH0ES at levels of $3.8\sigma$, $4.4\sigma$, and $4.3\sigma$ respectively. Our findings suggest that DESI BAO data favor sightly higher values of $H_0$ compared to the non-DESI BAO data employed in this study.





\textit{Acknowledgements.}
We acknowledge the use of HPC Cluster of ITP-CAS.  QGH is supported by the grants from NSFC (Grant No.~12475065, 11991052)  and China Manned Space Program through its Space Application System. XZ is supported by grant from NSFC (Grant No. 12005183).

\bibliography{refs}
\end{document}